# Low-temperature creep of binderless tungsten carbide with different grain sizes

E.A. Lantcev[1], A.V. Nokhrin[1(*)], V.N. Chuvil'deev[1], M.S. Boldin[1], Yu.V. Blagoveshchenskiy[2], P.V. Andreev[1], K.E. Smetanina[1], A.A. Murashov[1], N.V. Isaeva[2], A.V. Terentyev[2], N.Yu. Tabachkova[3,4]

[1] Lobachevsky State University of Nizhny Novgorod

[2] A.A. Baikov Institute of Metallurgy and Materials Science, Russian Academy of Science, Moscow, Russian Federation

[3] National University of Science and Technology "MISIS", Moscow, Russian Federation

[4] A.N. Prokhorov General Physics Institute, Russian Academy of Science, Moscow, Russian Federation

elancev@nifti.unn.ru

**Abstract**

The creep mechanism in the compression testing of the ceramic samples of binderless tungsten carbide with different grain sizes has been studied. The tungsten carbide samples with high relative density (96.1-99.2%) were obtained by Spark Plasma Sintering (SPS) from nano-, submicron, and micron-grade α-WC powders. In addition, ceramics sintered from submicron-grade powders with increased oxygen and graphite contents were studied. The ceramic samples had a nonuniform macrostructure with coarse-grained surface layers of ~300 μm in thickness and fine-grained central parts consisting of tungsten monocarbide α-WC with a small fraction of $W_2C$ particles. The creep tests were conducted in two regimes: (Mode 1) isothermic holding at different temperatures (1300-1375 °C) at constant stress (70 MPa); (Mode 2) tests at different stresses (50, 70, 90 MPa) at 1325 °C. Tests in Mode #1 were done to determine the effective creep activation energy $Q_{cr}$ while tests in Mode #2 – to determine the magnitude of coefficient *n* in the power law creep equation. The increasing of the volume fraction of the $W_2C$ particles from 1.7 up to 4% was found to result in a decrease in the creep activation energy of tungsten carbide $Q_{cr}$ from 17.5 down to 13 $kT_m$. The magnitude of coefficient *n* in the power law creep equation equals to 3.1-3.7. The creep activation energy in the ultrafine-grained (UFG) tungsten carbide with the grain sizes ~0.15 μm sintered from plasma chemically nanopowders was shown to be ~31 $kT_m$ (786 kJ/mol). This value is 1.5-2 times greater than the creep activation energy in the fine-grained tungsten carbide samples obtained by SPS from commercial powders. The increased volume fraction of the $W_2C$ particles formed when sintering the α-WC nanopowders with increased adsorbed oxygen concentration was suggested to be one of the origins of the increase in the creep activation energy $Q_{cr}$ when testing the UFG tungsten carbide samples. The mechanical removing of the coarse-grained layers from the surfaces of the tungsten carbide sample was shown to result in an accelerated creep, insufficient decrease in the activation energy and coefficient *n* to 2.5-2.6. The creep rate of the tungsten carbide samples obtained by SPS ($\dot{\varepsilon} = \dot{\varepsilon}_v + \dot{\varepsilon}_b$) was suggested to be determined simultaneously by the creep process in the coarse-grained surface layers ($\dot{\varepsilon}_v$) and the creep process in the fine-grained central parts of the samples ($\dot{\varepsilon}_b$). The creep rate in the surface coarse-grained layers ($\dot{\varepsilon}_v$) is determined by intensity of carbon diffusion in the α-WC crystal lattice while the creep rate in the fine-grained central parts ($\dot{\varepsilon}_b$) – by the intensity of grain boundary diffusion.

---

(*) corresponding author (nokhrin@nifti.unn.ru)

## 1. Introduction

At present, the products from tungsten carbide play a significant role in industry. The wear-resistance structural ceramics and hardmetals from tungsten carbide can be applied for making the ceramic cutting tool in machine building, drilling tools in mining, wear-resistance friction parts and wear components, mechanical seals, drawing dies, etc. [1-9]. Therefore, much attention in last decade was paid to the development of highly dense materials from binderless tungsten carbide having high strength, hardness, and fracture toughness [2, 10-16]. The forming of uniform nano- and fine-grained microstructure with increased density is one of the most efficient methods of improving the mechanical properties and performance characteristics of binderless tungsten carbide and WC-Co hardmetals [10, 12, 14, 16-25].

The structural ceramics from tungsten carbide should be operated at elevated temperatures and increased stresses. Therefore, the problem of investigation of the high-temperature deformation of tungsten carbide is relevant. At present, there are many studies devoted to the problem of room-temperature and at high-temperature deformation of coarse-grained binderless tungsten carbide and hard alloys based on tungsten carbide (WC-Co, WC-Fe etc.) [21, 27-49]. However, only few data on binderless tungsten carbide are available to date, and these are hard to explain within the fames of conventional models of high-temperature deformation. For example, Prof. S. Lay with co-workers [29, 30] have found the creep of binderless tungsten carbide to be featured by abnormally low values of activation energy $Q_{cr}$: 240 kJ/mol (60 kcal/mol), which are 2 times smaller than the creep activation energy in hard alloys WC-Co (~130 kcal/mol). Note that the values of the creep activation energy $Q_{cr}$ found in [29] were smaller that the activation energy of grain boundary diffusion of carbon $^{14}C$ in tungsten monocarbide (284 kJ/mol) [50] that is also an extremely unexpected result. Usually, the creep activation energy is suggested to be close to the activation energy of the dominating diffusion mechanism or appears to be greater than the one [46, 47, 51]. The magnitude of the coefficient $n$ in the power law creep equation for the binderless tungsten carbide samples varies from 2 to 3 [29, 30]. It is also interesting to note that the fine-grained ceramics from binderless tungsten carbide manifested the creep effect at relatively moderate stresses and low temperatures (1080-1380 °C ~ (0.44-0.54)$T_m$ where $T_m$ = 3053 K is the melting point of WC). One can suggest the creep in the fine-grained materials at such low temperatures to be provided by the grain boundary diffusion. However, the calculated vales of $Q_{cr}$ and $n$ don't match to the data available for Coble creep [52, 53]. In spite of a number of unexpected results, the works by Prof. S. Lay with co-workers [29, 30] remain the only studies of creep process in binderless tungsten carbide up to date.

The applications related to the process of Spark Plasma Sintering (SPS) of tungsten carbide are important scientific aspects of investigation of the creep mechanisms in tungsten carbide. SPS technology is a method of high-speed hot pressing, which allows implementing high heating rates (up to 2500 °C/min) with simultaneous application of pressure [10-15, 54-58]. Sintering is performed in vacuum or in an inert ambient by passing the millisecond high-power pulses of electric current that allows controlling the heating rate efficiently. Earlier, the creep intensity was shown to determine the compaction kinetics of the nano- and submicron-grade WC powders at the stage of intensive shrinkage [59, 60]. In SPS of the tungsten carbide nanopowders, the intensive shrinkage takes place at the stress of 30-80 MPa $\sim(1.1-2.9)\cdot 10^{-4}G$ in the temperature range 800-1400 °C $\sim (0.35-0.55)T_m$ [10-16, 21, 56, 59, 60], where $G = 275$ GPa is the shear modulus of WC, $T_m = 3053$ K is the melting point of WC. These temperature and stress ranges are close to the ones, which the creep in binderless tungsten carbide is observed in [29, 30]. Although the theory of creep in the sintering of porous materials is well developed [53, 61, 62], it is difficult to explain the origin of accelerated creep in this high-strength refractory ceramic material at such low stresses and temperatures. In our opinion, the abnormally low SPS temperature of the nano– and submicron-grade WC powders is partly a consequence of low temperatures, which the creep process of tungsten carbide takes place at.

Note that the objects of investigations in [29] were the tungsten carbide samples with increased porosity (~5%), which originated from the features of the fabrication technology. The possibility of obtaining the microstructure with small grain sizes, high relative density, and, as a consequence, with good mechanical properties in the structural ceramic based on binderless tungsten carbide is an important advantage of SPS technology [10-15, 56, 59, 63-65]. In our opinion, the application of SPS technology will help overcoming the problem of lower wear resistance and durability when cutting the cutting tools from binderless tungsten carbide as compared to commercial WC-Co hard alloys in the nearest future. At present, there are no data in the literature on the creep mechanisms in the fine-grained ceramics based on tungsten carbide fabricated by SPS.

The present work was aimed at the investigation of the deformation behavior of the tungsten carbide ceramic with different grain sizes and high relative density.

## 2. Materials and methods

The tungsten monocarbide α-WC powders with different initial particle sizes ($R_0$) were the objects of investigations: nanopowder #1 with initial particle sizes $R_0 \sim 95$ nm and commercial powders #2 and #3 manufactured by Kirovograd Hard Alloy Plant, JSC (Russia) with the initial particle sizes 0.8 μm and 3 μm (according to Fisher), respectively. Nanopowder #1 was obtained by DC arc plasma chemical synthesis followed by annealing in hydrogen. The technology of DC arc plasma chemical synthesis of tungsten carbide nanopowders was described in details in [66-69]. The characteristics of the investigated powders are presented in Table 1.

Table 1. Characteristics of the powders and ceramics

| Powder # | | 1 | 2 | 3 | 2.1[(1)] | 2.2[(2)] |
|---|---|---|---|---|---|---|
| Characteristics of the powders | | | | | | |
| Initial particle sizes $R_0$, nm | | 95 | 800 | 3000 | 800 | 800 |
| XRD phase analysis results, % | α-WC | 100 | 98 | 98 | 98 | 98 |
| | $W_2C$ (± 0.5) | - | 2 | 2 | 2 | 2 |
| Specific surface area, m²/g (±0.01) | | 4.05 | 0.48 | 0.13 | 0.48 | 0.48 |
| Carbon concentration, wt.% (±0.01) | | 6.31 | 6.14 | 6.13 | 6… | +0.1%C |
| Oxygen concentration, wt.% (±0.01) | | 0.30 | 0.15 | 0.12 | - | - |
| Characteristics of ceramics | | | | | | |
| Sintering temperature $T_s$, °C | | 1500 | 1590 | 1690 | 1650 | 1650 |
| Relative density, % (±0.05) | | 99.2 | 96.6 | 96.1 | 98.6 | 99 |
| Mass fraction of $W_2C$ particles, wt.% | | 5.9 ± 0.8 | 2.4 ± 0.5 | 2.1 ± 0.5 | 4 ± 0.5 | 1.8 ± 0.5 |
| Grain sizes, μm | | ~0.15 | ~1 | ~3 | ~1 | ~1 |
| $H_v$, GPa (±1) | | 27.7 | 24.3 | 21.0 | 24.1 | 23.7 |
| $K_{IC}$, MPa·m$^{1/2}$ (±0.3) | | 4.3 | 3.9 | 3.9 | 4 | 4.2 |

| | | | | | |
|---|---|---|---|---|---|
| Creep activation energy $Q_{cr}$, $kT_m$ (kJ/mol) | 31 (790) | 15 (380) | 22 (560) | 13 (330) | 17.5 (444) |
| Coefficient n (±0.5) | 3.1 | 3.7 | 2.4 | 5.2 | 3.7 |

(1) powder #2 after storing in air for 180 days

(2) powder #2.1 with 0.1% graphite added

To analyze the effect of the oxygen concentration and of the $W_2C$ particles on the creep processes in tungsten carbide, submicron powder #2 was stored in air for 180 days at room temperature. Mean humidity in the room, which the powder #2 was stored in was 40%. To analyze the effect of excess carbon of the creep kinetics, 0.1 wt.% of colloid graphite was added to the powder #2.1 after storing in air for 180 days. Graphite was added by mixing the WC powders with 0.1% carbon in isopropanol ambient. The graphite and WC powders were mixed with Pulverisette® 6 planetary mill. To reduce the pollution of external impurities, the lining of the milling cup and the milling balls were made from a hard alloy WC-Co. The powders were mixed for 6 hrs with the speed 150 with the 1-min pauses every hour to change the rotation direction. Prior to mixing, the WC + 0.1%C suspensions were subjected to homogenization with Hielscher® UP200Ht ultrasonic processor for 10 min. The homogenization was performed in isopropanol ambient. Residual ethanol was removed in 3 hrs using Binder® vacuum dryer box at 50℃ and the pressure 50 mbar.

The mean particle sizes in the powders ($R_{BET}$) were calculated from the specific surface area data ($S_{BET}$) measured by BET method: $R_{BET} = 6/(\rho_{th} \cdot S_{BET})$ where $\rho_{th}$ = 15.77 g/cm$^3$ is the theoretical density of tungsten monocarbide. The specific surface area was measured with Micrometrics® TriStar™ 3000 analyzer. The concentrations of oxygen and carbon were measured with Leco® CS-600 instrument with the uncertainty of ±0.01 wt.%.

The cylindrical tungsten carbide samples with the diameters D = 12 mm and the heights h = 12 mm were sintered using Dr. Sinter™ model SPS-625 (Japan) setup. The samples were sintered from the pressed powder workpieces of 22.5 g in weight in the continuous heating regime with the rate $V_h = 50°C/min$ up to the sintering temperature $T_s$. The ceramic samples were cooled down together with the setup. Holding at the sintering temperature $T_s$ was absent ($t_s = 0$ s). Sintering was performed in the conditions of uniaxial stress $\sigma = 70$ MPa, which was applied to the samples simultaneously with the start of heating. The temperature was determined by CHINO® IR-AH optical pyrometer focused onto the graphite mold surface. To improve the contact between the samples and the inner surface of the graphite mold, a graphite foil, was placed between the outer sample surfaces and the inner graphite mold one. The uncertainty of the temperature measurements was ±20 °C. In the course of experiment, the magnitude of effective shrinkage of the powders $L_{eff}$ was measured using the built-in dilatometer of Dr. Sinter™ model SPS-625 setup.

The creep testing of the sintered WC ceramic samples was conducted using Dr. Sinter™ model SPS-625. For the creep tests, the samples were placed into the graphite mold with the inner diameter 20 mm, which was greater than the one of the sintered ceramic samples (Ø12 mm). Prior to the testing, the tungsten carbide sample surfaces were subjected to waterjet treatment to remove the residual graphite. The creep tests were conducted in two regimes (Regimes #1, 2).

For testing in Regime #1, the samples were heated up with the rate $V_h = 25$ °C/min up to a preset temperature $T_{h1}$ and were kept at $T_{h1}$ for 30 min. The tests were conducted at a constant uniaxial stress $\sigma = 70$ MPa applied. In the course of experiment, the dependencies of the effective shrinkage on the isothermic holding time $L_{eff}(t)$ at different temperatures $T_{h1} = 1300, 1325, 1350$, and $1375$ °C were measured. The tests in Regime #1 were used to determine the effective creep activation energy $Q_{cr}$.

In the tests in Regime #2, the samples were heated up to the temperature $T_{h2} = 1325°C$ with the rate 25°C/min. Afterwards, the samples were kept at the stress $\sigma = 50$ MPa for 15 min (the first testing step). After holding for 15 min at $\sigma = 50$ MPa, the stress was raised up to 70 MPa, and the samples were kept at this pressure and $T = 1325°C$ for 15 min (the second test step). At the third test step, the stress was raised up to 90 MPa, and the samples were kept at $T = 1325$ °C for 15 min. In the course of experiments, the dependencies of the effective shrinkage on the time of experiment at different values of the stress applied were measured. The nominal stress values (50, 70, and 90 MPa) were recalculated into the actual ones on the base of the data on the change of the cross-section area of the cylindrical samples after the compression tests. The tests in Regime #2 were used to determine the magnitude of coefficient *n* in the creep equation.

To account for the contribution of the thermal expansion of the graphite mold, an experiment on the heating of empty molds was performed. True shrinkage (L) was determined by subtraction of the thermal expansion of the mold from the $L_{eff}(T, t)$ measured experimentally. The strain ($\varepsilon$, %) of the samples was calculated from the changes of shrinkage magnitudes: $\varepsilon(t,T) = (L(t,T) – L_0)/L_0$ where $L_0$ is the initial sample height.

The microstructure of the samples was investigated with Jeol® JSM-6490 Scanning Electron Microscope (SEM) and Jeol® JEM-2100F Transmission Electron Microscope (TEM). X-Ray Diffraction (XRD) phase analysis was carried out using Shimadzu® XRD-7000 diffractometer (CuK$_\alpha$, the scan step 0.04º, exposure time 2 s at every point). The qualitative phase analysis was carried out using Diffrac.EVA software. The quantitative analysis was carried out by Rietveld method. The uncertainties of determining the volume fractions of the α-WC particles and of the $W_2C$ ones were ±0.5%. The technique of the XRD phase analysis of the tungsten carbide samples is described elsewhere [59, 70]. The initial parameters of the phases were taken from ICSD compound database and PDF-2 powder diffraction database.

The Vickers hardness (Hv) of the ceramics was measured using Qness® A60+ microhardness tester. The measurements were performed at a load of 10 kg, the optimal loading time was 15 s. The minimal fracture toughness coefficient ($K_{IC}$) were calculated according to Palmquist model from the lengths of the longest radial cracks. When calculating $K_{IC}$, the elastic modulus magnitude for tungsten carbide was accepted to be equal to E = 710 GPa.

Prior to the investigations of the phase composition, microstructure, and mechanical properties, the sample surfaces after SPS were subjected to mechanical grinding and polishing to remove the carbonized layer of ~300-350 μm in thickness (see [59, 60]). The mechanical treatment was used to remove the abnormally large grains located at the sample sides because of intensive diffusion of carbon from the graphite mold (Fig. 1). Due to loose fit of the graphite foil to the sample surface, the carbonized layer depth for different parts of the sample may vary (see Fig. A2 in [59]). The intensive carbon diffusion is one of the origins of the modification of the properties of the surface layers of the metallic and ceramic samples obtained by SPS [59, 60, 70-78].

## 3. Results

3.1 Sintering of the ceramics. Making the samples for the creep tests

Fig. 2 presents the results of XRD phase analysis of the tungsten carbide powders studied. One can see from the XRD curves presented that the plasma chemically nanopowder #1 consists of the hexagonal tungsten monocarbide phase $\alpha$-WC (PDF 00-051-0939, ICSD 43380) completely. In the commercial powders #2 and #3, the tungsten carbide $W_2C$ phase (PDF 00-035-0776, ICSD 159904) was present. The mass fraction of the $W_2C$ particles in the powders #2 and #3 was ~2%. The XRD peaks from the nanopowder were broadened more as compared to the ones from the commercial powders #2 and #3 and were shifted towards greater diffraction angles insufficiently.

Storing the submicron-grade WC powders in air for 180 days resulted in an increase in the oxygen concentration in the powders from 0.15 up to 0.26 wt.%. Storing the powders #2.1 in air for 180 days didn't affect the XRD curves essentially. The XRD peaks corresponding to different modifications of tungsten oxide, which may form due to some chemical interactions of WC with oxygen (see [1, 59]) were absent in the XRD curves of the powder #2.1.

Figures 3 and 4 present the SEM and TEM image of the powders studied. The plasma chemically nanopowders #1 in the initial state consisted of large agglomerates up to 30-50 μm in diameter (Fig. 3a) consisting of separate nanoparticles (Fig. 3b). The homogenization and mixing in the planetary mill resulted in a destruction of the agglomerates. There were the nanometer-thick amorphous tungsten oxide layers on the WC nanopowder particle surfaces (Fig. 3c), which likely promote the strong tendency to agglomeration of the plasma chemically nanopowders WC. Twins formed as a result of an ultrafast crystallization were observed inside the WC nanoparticles (Fig. 3d).

No large agglomerates were observed in the commercial powders #2 and #3. The shapes of the particles were close to the spherical ones (Fig. 4a, b). The mean tungsten carbide particle sizes determined by chord method were close to the mean particle sizes determined according to Fisher. In the fine-grained powder #3, a considerable fraction of particles with mean sizes of 1-2 μm was observed (Fig. 4b). It evidences a broad enough size distribution of the particles.

Fig. 5a presents the temperature curves of the shrinkage L(T) and of the shrinkage rate S(T) of the powders #1-3. One can see the curves L(T) to have conventional three-stage character (see [53]): an insufficient compaction at low temperatures 600-1000 °C (Stage I), an intensive shrinkage in the heating temperature range of 1000-1400 °C (Stage II). In the temperature range from 1400 °C to the sintering temperature $T_s$ (Stage III), the powder shrinkage intensity decreases again. The decreasing of the initial particle sizes resulted in a decrease in the optimal sintering temperature $T_s$ – for the powders of Series #1 ($R_{01}$ ~ 0.1 μm), the sintering temperature $T_s$ was 1500 °C while for the powders from Series #2 ($R_{02}$ ~ 0.8 μm) and #3 ($R_{03}$ ~ 3 μm), the values of $T_s$ were 1590 °C and 1690 °C, respectively. One can se in Fig. 5a the maximum values of the shrinkage ($L_{max}$) and of the shrinkage rate ($S_{max}$) to decrease with increasing initial particle sizes. For the nanopowder #1, the magnitudes $L_{max}$ and $S_{max}$ were ~25 mm and 0.04 mm/s, respectively. For the powders #2 and #3, the values of $L_{max}$ were ~17 and ~13 mm respectively, while the maximum shrinkage rates $S_{max}$ were ~ 0.03 and 0.02 mm/s, respectively. The temperature $T_{max}$, which the maximum value of the shrinkage rate $S_{max}$ was achieved at ($T_{max} = T (S = S_{max})$) decreased with increasing $R_0$ – for the nanopowders #1, #2, and #3, the magnitudes of $T_{max}$ were 1020-1080 °C, 1370-1470 °C, and 1510 °C, respectively (Fig. 5a).

Fig. 5b presents the temperature curves of the shrinkage and of the shrinkage rate for the powders #2, #2.1, and #2.2 differing from each other by the concentrations of oxygen and carbon (Table 1). As one can see from the curves L(T) and S(T) presented in Fig. 5b, the sintering kinetics at Stage I and Stage II for the powder #2.1 with increased oxygen contents and the powder #2.2 with the addition of graphite were similar to the compaction kinetics for the submicron-grade powder #2. The maximum shrinkage rate $S_{max}$ for the powder #2 was $3.5 \cdot 10^{-2}$ mm/s, and for the powders #2.1 and #2.2 $S_{max}$ appeared to be close to $3.0 \cdot 10^{-2}$ mm/s. The maximum shrinkage $L_{max}$ for the powder #2 at $T_s$ was 16.8 mm whereas the samples from the powders #2.1 and 2.2 demonstrated close values of $L_{max}$ equal to 13.8 and 13.5 mm, respectively.

The relative density of the ceramics was high and was close to 100% for the samples of Series #1 and 98% for the samples of Series #2, 3 (Table 1). For convenience, here and hereafter the ceramic samples sintered from the powders #1-3 will be marked by the same numbers as the powders, which these one were made form.

Fig. 6 presents the photograph of the microstructure of the sintered tungsten carbide samples. The analysis of the electron microscopy investigation results shows the ceramics #1 to have a uniform UFG microstructure with mean grain sizes ~100-200 nm (Fig. 6a). The mean grain size (d) in the ceramics #2 and #3 were ~1 and ~3 μm, respectively (Fig. 6c, 6e). There were no abnormally large grains in the sintered ceramics. The pores located preferentially at the grain boundaries were present in the microstructure of the ceramics. The mean pore sizes increased with increasing mean grain sizes in the ceramics (Fig. 6).

The photographs of the microstructure of the samples #2.1 and #2.2 presented in Fig. 7 show the additional adsorption of 0.10-0.11 wt.% oxygen (#2.1) or additional of 0.1 wt.% carbon (#2.2) not to result in a notable change in the character and parameters of the microstructure of the sintered tungsten carbide samples. It is worth noting that the samples sintered from the powder #2.2 were featured by smaller numbers of pores (the black areas in the photographs) and, as a consequence, by somewhat greater values of relative density (Table 1). This result agrees well with the data known from the literature that the additional introduction of carbon into the tungsten carbide powders allows increasing the density of the ceramics [59, 60, 78-80]. Besides, the samples #2.2 had somewhat larger grain sizes than the samples #2.1 (Fig. 7a, c).

Fig. 2 presents the results of the XRD phase analysis of the sintered ceramics. The $W_2C$ phase was present in all ceramic samples. Analysis of the XRD phase analysis results shows a lower $W_2C$ content to be observed in the ceramics sintered from the factory made powders as compared to the ceramics #1. Thus, the $W_2C$ phase content in the ceramic samples #1 was 5.9 ± 0.8 wt.% whereas the mass fractions of $W_2C$ in the samples #2 and #3 were 2.4 ± 0.5 and 2.1 ± 0.5%, respectively[1]. As one can see from Table 1, the contents of the $W_2C$ particles in the ceramics #2 and #3 were close to the ones of the $W_2C$ phase in the initial powders #2 and #3. No XRD peaks corresponding to the α-W phase (PDF 00-004-0806, ICSD 76151) were observed (Fig. 2). So far, no increasing of the contents of $W_2C$ phase particles took place during SPS of the commercial powders #2 and #3 whereas an essential increase in the content of $W_2C$ particles was observed during sintering the plasma chemically synthesized. In our opinion, this is related to the increased oxygen content in the plasma chemically synthesized WC nanopowders (Table 1).

One can see from Table 1 that the increasing of the amount of adsorbed oxygen in the powder #2 resulted in an increase in the amount of the $W_2C$ phase in the ceramics from 2.4% (for the ceramic #2) up to 4% (for the ceramic #2.1). The addition of 0.1 wt %C into the powder #2.1 reduced the amount of the $W_2C$ phase in the sintered ceramics down to 1.8% (for the ceramic #2.2).

Table 1 presents the results of the microhardness $H_v$ and the minimal fracture toughness coefficient $K_{IC}$ measurements in the sintered tungsten carbide samples. One can see the hardness of the ceramics to decrease with increasing mean grain sizes. Thus, the UFG tungsten carbide samples sintered from the nanopowder #1 (d = 100-200 nm) had the hardness of ~28 GPa whereas the samples #2 (d ~ 1 μm) and #3 (d ~ 3 μm) had the hardness values 24.2 and 21.0 GPa, respectively. The minimal fracture toughness coefficient values were somewhat smaller than the results presented in [63, 81] that is likely related to the presence of the $W_2C$ phase in the sintered samples.

---

[1] The mean mass fraction of the $W_2C$ particles was determined by averaging the data for 4 ceramic samples or more.

The increasing of the amount of oxygen in the powder #2 as well as the additional of carbon into the powder #2.1 didn't result in the change of the mechanical properties of the sintered samples (Table 1). Thus, the average values of hardness $H_v$ for the samples #2.1 and #2.2 were 24.1 and 23.7 GPa, respectively. The fracture toughness coefficient $K_{IC}$ for the ceramics #2.1 and #2.2 were 4.0 and 4.2 MPa·m$^{1/2}$, respectively. These values are close to the characteristics of the tungsten carbide samples sintered from the initial powder #2: $H_v$ = 24.3 GPa, $K_{IC}$ = 3.9 MPa·m$^{1/2}$ (Table 1).

3.2 Creep testing of the ceramics with different grain sizes

The samples #1-3 made within the previous stage of the work (see Subsection 3.1) were used for the creep tests. Prior to the testing, the sample surfaces were subjected to the water jet cleaning to remove the residual graphite. Cleaning was done using Gidroabraziv™ MS41 setup.

The isothermic holdings were conducted at 1300, 1325, and 1350 °C for ceramics #1 and #2, and at 1325, 1350, and 1375 °C for ceramic #3. The isothermic holding temperatures selected for testing in Regime #1 corresponded to the intensive powder shrinkage stage (Fig. 5).

Fig. 8a and 8c presents the dependencies of the strain of the samples on the isothermic holding time at the stress $\sigma$ = 70 MPa. The curves $\varepsilon(t)$ had three-stage character: the stage of the thermal expansion of the samples, which the effective strain of the samples can be negative at (Stage I), the initial stage of the creep, which rapid increase of the sample strain is observed at (Stage II), and, finally, the stage of steady-state creep (Stage III), which a linear character of the $\varepsilon(t)$ curve is observed at. The creep rates were calculated for the linear parts of the $\varepsilon(t)$ curves. The analysis of the $\varepsilon(t)$ curves presented in Fig. 8a and 8c shows the strain and the strain rate to increase with increasing holding temperature. The maximum strain $\varepsilon_{max}$ for the ceramics #1 and #2 increased from 5.8 up to 29.5% and from 6.7 up to 19%, respectively with increasing testing temperature from 1300 °C up to 1350 °C. The maximum strain of the ceramic #3 increased from 4.1 up to 18.5% with increasing isothermic holding temperature from 1325 °C up to 1375 °C. The dependencies of the maximum strain for the ceramics #1-3 on the testing temperature in the semi-logarithmic axes $\ln(\varepsilon_{max}) - T_m/T$ can be fitted by a straight line with good accuracy (Fig. 9). The reliability coefficient of the linear fit for the dependence $\ln(\varepsilon_{max}) - T_m/T$ $R^2 > 0.96$.

To study the effect of stress on the creep rate of tungsten carbide, the samples were tested at 1325 °C varying the stress applied from 50 up to 90 MPa with 15 min holding at each stress value. Fig. 8b, d present the curves strain - time at different stress values, the time dependencies of the force applied to the sample F(t) are presented also[2]. One can see a step-wise increase in the strain rate with increasing stress. Afterwards, the steady-state creep stage with a constant strain rate was observed. The increasing of the stress resulted in an increase in the strain rate of the ceramics. It is worth noting that an increase in the strain of the samples being tested was observed when decreasing the mean tungsten carbide grain sizes. Thus, the strain $\varepsilon_{max}$ for the ceramics #1, 2, and 3 in the end of the experiments were 49, 31.8, and 12%, respectively at the magnitude of the load 16.5 kN applied. So far, the decrease in the grain size leads to an increase in the maximum deformation of the ceramics $\varepsilon_{max}$.

High magnitude of strain of the UFG ceramic #1 was confirmed by the results of measurements of the geometrical sizes of the samples after the tests (Fig. 1b). One can see in Fig. 10 the tungsten carbide sample deformation in the creep regime to take place without the failure of the samples and without formation of macrodefects in the form of large pores or cracks. The deformation of the samples was uniform, without the loss of the shape stability.

---

[2] We have chosen the data presentation format with F(t) intentionally to show the raw data on the force F(t), which then were recalculated into the values of stress σ at each of three stages of testing according to the formula $\sigma_{1,2,3} = F_{1,2,3}/S_{1,2,3}$ where $F_{1,2,3}$ are the mean values of the force acting on the sample within the stationary flow stage (in kN) and $S_{1,2,3}$ are the mean values of the sample area at this stage measured experimentally. The procedure introduced resulted in an insufficient excess of the actual values used in the calculations over the nominal ones (50, 70, and 90 MPa) in 4-6 MPa.

The comparative analysis of the sample microstructure before and after creep tests (Fig. 6) shows the deformation at elevated temperatures not to result in changes in the shapes and sizes of the WC grains. Note also that the tungsten carbide grains keep their initial shape after the creep testing (Fig. 6). For the samples tested at 1350 ºC, the mean grain sizes in the central parts of the samples correspond to the ones at the sample edges. The mean sizes and the volume fraction of pores in the samples after the creep tests correspond to respective parameters for the samples in the initial state (before the creep tests) (Fig. 6). So far, one can conclude no considerable plastic growth of pores to be observed in the tungsten carbide samples during creep tests.

3.3 Creep testing of the ceramic samples after the mechanical treatment

In the course of SPS, the tungsten carbide sample surfaces are saturated with carbon, which diffuses into these ones from the graphite mold and the graphite foil (see also [57, 59, 60, 70-77]). This leads to the altering of the phase composition and the mechanical properties of the surface layers of the sintered tungsten carbide samples [59, 60]. As it is shown in Fig. 1, the diffusion of carbon resulted in an intensive growth of the tungsten carbide grains and in the formation of the coarse-grained microstructure in the surface layers. The presence of the abnormally large grains may affect the creep characteristics of the tungsten carbide samples.

To remove the coarse-grained surface layers, the sintered tungsten carbide samples were subjected to additional mechanical treatment with Buehler® Phoenix™ Beta grinding machine. The depths of the removed layers were 1.5 mm. The density check has shown the additional mechanical treatment not to affect the relative density of the tungsten carbide samples. The metallographic investigations revealed no defects in the form of large cracks on the sample surfaces after additional mechanical treatment. In Fig. 9 and thereafter, the ceramic samples #2.1 after the additional surface treatment are marked as Series 2.1[*].

Fig. 11 presents the curves strain - isothermic holding time for ceramic #2.1[*], which were subjected to the preliminary additional mechanical treatment. For the convenience of comparison, the

$\varepsilon(t)$ curves for the ceramics without the additional treatment are presented in Fig. 11 also. The removal of the surface layers resulted in a shift of the characteristic deformation temperatures of the ceramics in ~25 °C towards the lower temperatures as compared to the non-treaded ceramics. The maximum strain of the ceramic samples ($\varepsilon_{max}$) increased with increasing testing temperature. The dependence of the maximum strain of the samples on the testing temperature in the semi-logarithmic axes $\ln(\varepsilon_{max})$ – $T_m/T$ can be fitted by a straight line with good accuracy. It is important to note that the strain magnitude $\varepsilon_{max}$ for the treated samples appeared to be greater than for the non-treated ceramic samples #2.1. The differences in the values of $\varepsilon_{max}$ for the treated samples and non-treated ones increase with increasing test temperatures. The values of the maximum strain magnitude $\varepsilon_{max}$ for the initial ceramic samples and the treated ones #2.1 were 4.0 and 6.6 % at the testing temperature of 1300 °C and, 6.8 and 22% at the one of 1325 °C, respectively. So far, it should be stressed once more that the removing of the coarse-grained surface layers leads to the increase in deformation degree of the tungsten carbide samples.

The tungsten carbide samples of Series #2.1 subjected to the additional mechanical treatment preserved the uniform fine-grained microstructure after the creep tests (Fig. 12). The increasing of the deformation temperature didn't affect the character of pore distribution and the volume fraction of the pores in the ceramic samples #2.1$^*$ essentially (Fig. 12).

### 3.4 Effect of oxygen and carbon on the creep of the ceramics

To investigate the effect of carbon and oxygen on the creep behavior of tungsten carbide at elevated temperature, 0.1 wt.% graphite was added into the powder #2.1 with the initial particle size of 800 nm and increased oxygen concentration. One can see from Table 1 that storing the powder #2.1 in air resulted in an increase in the mass fraction of the $W_2C$ phase in the sintered ceramics from 2.5 up to 4%. The addition of 0.1% of graphite into the powder #2.1 resulted in a decrease of the $W_2C$ phase content in the sintered ceramics down to 1.8%. The creep tests in Regime #1 were carried out at 1300, 1325, and 1350 °C, respectively that corresponds to the intensive powder shrinkage stage

(Fig. 5b). Fig. 13 presents the samples ceramic #2.1 after testing in Regime #1 at different temperatures. One can see the maximum strain of the samples to increase with increasing holding temperature and magnitude of the pressure applied. The samples deformed uniformly, without fracture and formation of large macrodefects (cracks).

Fig. 14a, c presents the curves strain – holding time for the ceramics #2.1 and #2.2 at different temperatures. These curves had a three-stage character as well as the dependencies L(t) for the samples of series #2 (Fig. 8).

One can see from the comparison of Fig. 14a and Fig. 14c the addition of graphite into the powder #2.1 to reduce the degree of ceramic strain at the same holding temperatures. The maximum strain increased from 4 up to 13.5% for ceramic #2.1 and from 0.5 to 5.2% for the ceramic #2.2 with increasing testing temperature from 1300 to 1350 °C. The dependencies of the maximum strain on the testing temperature in the $\ln(\varepsilon_{max}) - T_m/T$ axes for the ceramics #2.1 and #2.2 can be fitted by straight lines with good accuracy (Fig. 9).

Fig. 14b, d presents the dependencies of strain and the magnitude of the load applied on the experiment time at 1325 °C. As in the testing in Regime #1, the adsorbed oxygen and the addition of graphite resulted in a decrease in the strain of the tungsten carbide samples. For the ceramic samples #2, #2.1, and #2.2, the magnitudes of $\varepsilon_{max}$ after completing the tests in Regime #1 at the load F = 16.5 kN were 31.8, 24.5, and 15.4%, respectively.

The comparative analysis of the microstructure parameters of the samples ceramics before and after the creep tests shows the deformation of tungsten carbide not to result in changes in the shapes and sizes of the pores. The shapes and sizes of the WC grains didn't change after the creep tests also (Fig. 7).

Summarizing the experimental results described in this section, one can conclude that an intensive enough deformation of the binderless tungsten carbide samples was observed in the low-temperature creep conditions. The strain magnitude of the ceramics depended on the grain sizes – the sample strain magnitude increased with decreasing grain sizes. The microstructure parameters of the ceramics and the porosity of these ones didn't change in the course of creep testing – the tungsten carbide grains preserved their regular equiaxial shapes, no intensive grain growth or crack formation were observed during the creep tests. So far, the deformation of ceramics observed is not related to the decrease in the volume fraction of pores or to the fracture of the samples. Note that the sintered samples had nonuniform macrostructure – the coarse-grained layers were observed on the surfaces of the ceramics arising from intensive diffusion of carbon from the graphite mold. Removing this coarse-grained layers resulted in increase in the strain magnitude of tungsten carbide. Also, adding oxygen and carbon into the submicron-grade WC powder was found to result to the decrease in the deformation degree of the ceramics.

## 4. Discussion

### 4.1 Determining the creep mechanism

The mechanisms of SPS of the tungsten carbide powders were described in details elsewhere [59, 60, 63, 82]; and we will not focus on these ones here. Note only that according to [59, 60, 82], the creep is one of main mechanisms of compaction of the nano- and submicron-grade tungsten carbide powders at the intensive compaction stage.

The power law creep equation can be written in the form [51, 53]:

$$\dot{\varepsilon} = AD_{eff}(Gb/kT)(\sigma/G)^n, \qquad (1)$$

$$D_{eff} = D_0 \exp(-Q_{eff}/kT), \qquad (2)$$

where A is a dimensionless coefficient; $D_{eff}$ is the effective diffusion coefficient, b is the Burgers vector, G = 275 GPa is the shear modulus, k is the Boltzmann constant, and $Q_{eff}$ is the effective activation energy for the creep process.

According to [51, 53], the creep mechanism can be determined tentatively on the base of the magnitude of coefficient *n* in the creep equation (1) and of the one of effective creep activation energy $Q_{cr}$ в (2). For example, if *n* = 3 and the magnitude of $Q_{cr}$ is close to the activation energy of the volume diffusion $Q_v$, the power-law creep is the dominant deformation mechanism [51, 53]. The case of *n* =

1 and the grain boundary diffusion activation energy $Q_{cr} \sim Q_b$ is related to Coble creep usually [52, 53].

The effective creep activation energy $Q_{cr}$ was determined on the base of analysis of the $\varepsilon(t)$ curves (Fig. 8a, c) according to the procedure described in [51, 53]. From the slopes of the linear dependencies $\varepsilon(t)$, the creep rate ($\dot\varepsilon$) was determined. The values of the creep rate ($\dot\varepsilon_1, \dot\varepsilon_2, \dot\varepsilon_3$) were calculated for all temperatures ($T_1$, $T_2$, and $T_3$), which the tungsten carbide samples were tested at. The creep activation energy $Q_{cr}$ at the isothermic shrinkage stage can be determined from the slope of the dependence $\ln(\dot\varepsilon)$ - $T_m/T$ (Fig. 15a) and the magnitude of the coefficient $n$ – from the slope of the dependence $\ln(\dot\varepsilon)$ - $\sigma/G$ (Fig. 15b) [51-53]:

The calculated values of the creep activation energy $Q_{cr}$ are presented in Table 1. The uncertainty of determining the activation energy $Q_{cr}$ was ±2 $kT_m$ (± 50 kJ/mol). The analysis of the results obtained shows the creep activation energy for the ceramic #1 was equal to ~31 ± 2 $kT_m$ (~790 ± 50 kJ/mol) while the ones for ceramics #2 and #3 were ~15 ± 2 $kT_m$ (380 ± 50 kJ/mol) and 22 ± 2 $kT_m$ (560 ± 50 kJ/mol), respectively. Note that the values of $Q_{cr}$ for the ceramics sintered from the powders #1 and #3 were greater than the activation energy of carbon $^{14}C$ diffusion in the crystal lattice of tungsten monocarbide ($Q_v \sim$ 360 kJ/mol [50]). The results obtained evidence no dependence of $Q_{cr}$ on the ceramic grain size observed. The maximum value of $Q_{cr}$ was observed in the ceramic #1 containing the highest amount of $W_2C$ particles (Table 1).

Let us determine the magnitude of the coefficient $n$ in Equation (1). According to [51, 53], the magnitude of coefficient $n$ depends on the dominating creep mechanism. The magnitude of $n$ was determined from the analysis of the dependencies of shrinkage on the stress applied. The magnitudes of the strain rate ($\dot\varepsilon$) were determined form the slopes of the linear parts of the curves $\varepsilon(t)$ for each stress value (50, 70, and 90 MPa). The magnitude of the coefficient $n$ was calculated from the slope of the dependence of the strain rate on the magnitude of stress applied in the logarithmic axes $\ln(\dot\varepsilon)$ - $\ln(\sigma/G)$. The analysis of the results demonstrated the magnitude of coefficient $n$ for the ceramics #1-3 to decrease insufficiently from 3-3.7 to 2.4 with increasing mean ceramic grain size (Fig. 15b). The uncertainty of determining the magnitude of $n$ was ±0.4. Taking into account the uncertainty of determining the magnitude of $n$, one can conclude that no essential differences in the magnitudes of $n$ for the ceramics #1-3 were observed. The values of coefficient $n$ obtained evidence the power-law creep to be the primary mechanism of deformation of the fine-grained tungsten carbide in the ranges of temperatures (1300-1375 °C) and stresses (50-90 MPa) investigated. The values of coefficient $n$ determined in our experiment were close enough to the values reported in [29, 30].

### 4.2 Effect of oxygen and carbon

Analysis of the data presented in Section 3 shows the increasing of the oxygen and carbon contents in the submicron-grade WC powder leads to the decrease in the maximum deformation ($\varepsilon_{max}$) of the ceramic #2.

It follows from Table 1, that the increasing of the oxygen concentration in the powder resulted in an increase in the $W_2C$ phase content in the ceramic #2 while the increasing of graphite content– in an decrease of $W_2C$ content in the ceramic #№2. This result agrees well with the data reported in [1, 59, 60, 79, 81, 83-87]. Analyzing the data presented in Fig. 16, one can conclude the increasing of the amount of the $W_2C$ phase in the ceramic #2 from 2.5 to 4% to result in an insufficient decrease in the creep activation energy $Q_{cr}$ from 15 to 13 $kT_m$ that is comparable to the uncertainty of determining the magnitude of $Q_{cr}$. The addition of graphite reduces the $W_2C$ phase content in the sintered ceramics from 4.0 down to 1.8% while the creep activation energy $Q_{cr}$ increased from 13.0 up to 17.5 $kT_m$. The mean sizes of the tungsten carbide grains changed insufficiently (Table 1). The result obtained allows concluding the increasing of $W_2C$ content from 1.8 to 4% at fixed grain sizes $d \sim 1$ μm to result in a decrease in the creep activation energy in tungsten carbide $Q_{cr}$ from 17.5 down to 13 $kT_m$.

The magnitude of coefficient $n$ for the ceramic #2.2 is 3.7. So far, one can conclude the variation of the contents of the $W_2C$ particles in the sintered ceramics not to result in change of the creep mechanism of tungsten carbide but to affect the magnitude of the effective creep activation energy $Q_{cr}$ only.

### 4.3 Effect of the surface treatment

Removing the surface layers of the ceramics #2.1 containing the abnormally large grains resulted in insufficient decrease in the effective creep activation energy down to $Q_{cr} = 12 \pm 2$ $kT_m$. The value of $Q_{cr} \sim 12$ $kT_m$ (304 kJ/mol) obtained is close to the activation energy of grain boundary diffusion of carbon $^{14}C$ in tungsten monocarbide $Q_b = 11.7$ $kT_m$ (296 kJ/mol [50]). The value of coefficient $n$ for the samples tested was ~2.5-2.6. This value matches to the one obtained for the ceramics #2 and #2.2. So far, the removing of the surface layers of the abnormal grains leads to the decrease in the magnitude of coefficient $n$ but doesn't lead to essential change in the effective creep activation energy. Note also that, as one can see in Fig. 9, the removing of the coarse-grained surface layers leads to essential increase in the maximum strain of the ceramic ($\varepsilon_{max}$).

### 4.4 Summary of the results

The analysis of the results presented in subsections 4.1-4.3 shows no clear dependence of the creep activation energy $Q_{cr}$ on the $W_2C$ content found (Fig. 1b). At the same time, it is important to pay attention to the following key results:

(i) at constant grain sizes (d ~ 1 μm), the increasing of the $W_2C$ particle content from 1.8 up to 4% results in insufficient decrease in the effective creep activation energy, which is comparable to the uncertainty of determining the magnitude $Q_{cr}$ ($\pm 2$ $kT_m$).

(ii) the ceramics with close $W_2C$ contents having different mean grain sizes have essentially different activation energies $Q_{cr}$. This is quite unexpected result since usually the power law creep rate supposed not to depend on the grain sizes (see Equation (1)).

(iii) removing the coarse-grained surface layers by mechanical treatment doesn't lead to an essential change of the effective creep activation energy ($Q_{cr} \sim 12$ $kT_m$) but is accompanied with decrease in the coefficient n down to ~ 2.5-2.6. The content of the W$_2$C particles remained constant (3.5-4%W$_2$C).

So far, one can conclude the grain sizes to affect the creep kinetic more essentially than the contents of the W$_2$C particles.

Also, it is important to take into account that the tungsten carbide samples after SPS have nonuniform macrostructure. There are coarse-grained tungsten monocarbide layers up to 0.25-0.3 mm in thickness with increased carbon contents on the surfaces of the samples. So far, the coarse-grained layer area can achieve ~ 5% of the total surface of the sample of 12 mm in diameter. The W$_2$C particles in the surface tungsten monocarbide layers are almost absent [59]. The central (main) parts of the samples have uniform fine-grained microstructure with increased contents of W$_2$C particles. There are large enough pores located at the grain boundaries.

The creep rate of the tungsten carbide samples made by SPS at low temperatures (T ~ 0.5T$_m$) and at small stresses ($\sigma/G$ ~ $10^{-4}$) can be represented in the form of a sum:

$$\dot{\varepsilon} = \dot{\varepsilon}_b + \dot{\varepsilon}_v, \quad (3)$$

where $\dot{\varepsilon}_b$ is the grain boundary deformation rate and $\dot{\varepsilon}_v$ is the power law creep rate, the intensity of which is limited by the volume diffusion. In the first approximation, the grain boundary deformation rate can be accepted to be equal to the Coble creep rate [52, 53]:

$$\dot{\varepsilon}_{b1} = A_{b1}(\delta D_b/d^3)(G\Omega/kT)(\sigma/G). \quad (4)$$

where $A_b$ = 47.5 is a constant, $\delta$ is the grain boundary width, $\Omega$ is the atomic volume, and $D_b$ is the grain boundary diffusion coefficient.

In some cases, the grain boundary strain rate in the fine-grained ceramics can be accepted to be equal to the grain boundary sliding (GBS) rate [88, 89]:

$$\dot{\varepsilon}_{b2} = A_{b2}(\sigma/G)^2(b/d)^2(G\Omega/kT)(\delta D_b/b^3), \quad (5)$$

where $A_{b2}$ ~100 is a numerical coefficient [64].

According to [51, 53], the contribution of $\dot{\varepsilon}_v$ can be written in the following form:

$$\dot{\varepsilon}_v = A_D(\sigma/G)^3(G\Omega/kT)(D_v/b^2), \quad (6)$$

where $A_D$ is the Dorn constant, $D_v$ is the diffusion coefficient in the crystal lattice.

Form comparing the formulae (4)-(6), one can suggest the creep process in the fine-grained parts of the ceramic samples to be determined by the grain boundary strain rate $\dot{\varepsilon}_b$ while the creep process in the coarse-grained material of the surface layer – by the power law creep equation (6). It is important to stress that the grain boundary strain rate will be greater than the power law creep one ($\dot{\varepsilon}_b > \dot{\varepsilon}_v$) in the same deformation modes (T ~ 0.5T$_m$, σ/G ~ $10^4$). It originate from a greater values of the grain boundary diffusion coefficient $D_b$ as compared to $D_v$ as well as from small grain sizes $d$.

According to [51, 53], the process with the lowest strain rate will determine the resulting (effective) strain rate of the material as a whole. One can suggest that the effective strain rate in this case will be determined by the power law creep rate $\dot{\varepsilon}_v$ depending, first of all, on the diffusion coefficient in the crystal lattice $D_v$. The value of coefficient $n$ in this case is close to the theoretical one $n = 3$ [51].

In the case of removing the surface coarse-grained layers, the effective strain rate of the ceramic samples will be determined by the grain boundary strain one: $\dot{\varepsilon} \sim \dot{\varepsilon}_b$. In this case, one can expect a decrease in the coefficient $n$ down to ~ 1 in the case of Coble creep (see. (4)) or down to $n$ ~2 in the case of GBS (see (5)). The creep activation energy in this case should be close to the grain boundary diffusion ($Q_{cr} \sim Q_b$). Summarizing the results presented in subsection 4.3 allows suggesting the GBS to be main deformation mechanism in the fine-grained central parts of the tungsten carbide samples.

Finally, it is necessary to discuss the origins of the abnormal increase in the activation energy $Q_{cr}$ in the ultrafine-grained ceramics sintered from the plasma chemically synthesized WC nanopowders. As one can see from Fig. 16, in spite of considerable content of the $W_2C$ particles leading to the decrease in $Q_{cr}$, the magnitude of $Q_{cr}$ in the ceramics #1 reaches ~ 32 kT$_m$ (811 kJ/mol). This is an abnormally large value, which exceeds the activation energy of the volume diffusion ($Q_v$ ~ 360 kJ/mol [50]) and the one of grain boundary diffusion ($Q_b$ ~ 284 kJ/mol [50]) in tungsten carbide essentially.

One can see from Fig. 3b, 3c that the amorphous tungsten oxide layers are present on the surfaces of the tungsten carbide nanoparticles. Tungsten oxide, which interacts with carbon from the crystal WC lattice when heating that leads to the formation of $W_2C$ [1, 85, 86]. Carbide $W_2C$ can form also in the interaction of tungsten oxide with tungsten monocarbide [1]. In our opinion, The $W_2C$ nanoparticles forming at the grain boundaries prevent the grain boundary sliding and, therefore, lead to the increase in the activation energy $Q_{cr}$. The effect of the $W_2C$ nanoparticles on the motion of dislocations in the coarse-grained surface tungsten carbide layers can be the second factor promoting the increasing of the effective creep activation energy in ceramics #1. Initially, the $W_2C$ nanoparticles form on the surfaces of the plasma chemically synthesized tungsten carbide nanoparticles. In the case of intensive migration of the grain boundaries, the $W_2C$ boundaries will be located inside the WC grains. According to [51], the dispersed particles will prevent the motion of the lattice dislocations to overcome the obstacles of this type, some additional energy is needed. This will lead to the increase in the activation energy $Q_{cr}$.

Note that the suggestion made on simultaneous grain boundary diffusion and the volume one (see Equation (3)) allows explaining the contradiction arising during the analysis of the experimental data. As one can see in Fig. 9-10, the ceramic #1 after the creep tests had the largest strain magnitude, which exceeds the strain value $\varepsilon_{max}$ for the ceramics #2 and #3 essentially. At the same time, the analysis of experimental results presented in Subsection 4.1 shows the ceramic samples #1 to have ~ 2 times greater effective creep activation energy $Q_{cr}$. In our opinion, this contradiction is related to the dependence of the grain boundary strain rate on the grain sizes: $\dot{\varepsilon}_{b1} \sim A_1/d^3$ in the case of Coble creep (see (4)) in the case of GBS $\dot{\varepsilon}_{b2} \sim A_2/d^2$ (see (5)). The decreasing of the mean grain sizes from ~1-3 μm for the ceramics #2-3 down to ~0.1-0.2 μm for the ceramic #1 will lead to an essential increase in the creep rate and, hence, to the increase in the strain magnitude $\varepsilon_{max}$. Using Equation (5), one can show that the increasing of the activation energy $Q_{cr}$ from 15 up to 32 $kT_m$ with simultaneous decreasing of the mean grain size $d$ from 1 down to 0.1 μm will lead to the increase in the deformation rate 3 times.

**Conclusions**

1. The samples of 12 mm in diameter and 12 mm in height for the creep tests were made from the WC nano-, submicron-grade, and fine powders by SPS. The tungsten carbide samples obtained by SPS have a nonuniform macrostructure – there are the ~0.3 mm thick tungsten monocarbide layers with coarse-grained microstructure on the surfaces on the samples while a uniform fine-grained microstructure with increased fraction of $W_2C$ particles was observed in the central parts of the samples. The relative density of the samples scatters from 98% to 100%.

2. The effective creep activation energy $Q_{cr}$ for the samples sintered from the fine-grained commercial powders decreases insufficiently with increasing content of $W_2C$ particles. The magnitude of coefficient *n* in the power law creep equation is 3.1-3.7. The activation energy $Q_{cr}$ is close to the activation energy of carbon diffusion in the crystal lattice of tungsten carbide. The creep characteristics of the tungsten carbide samples are suggested to be determined by the characteristics of the coarse-grained surface layers.

3. After mechanical treatment removing the coarse-grained surface layers, the magnitude of coefficient *n* decreases down to 2.5-2.6 while the activation energy $Q_{cr}$ decreases down to ~ 12 $kT_m$. The grain boundary sliding (GBS) is suggested to be the primary deformation mechanism for the fine-grained tungsten carbide.

4. The creep activation energy in the ultrafine-grained (UFG) tungsten carbide with the grain sizes ~0.15 μm sintered from the plasma chemically synthesized nanopowders is ~31 $kT_m$. It is 1.5-2 times greater than the creep activation energy in the fine-grained tungsten carbide samples obtained by SPS from the commercial powders. The increased fraction of the $W_2C$ particles forming during sintering the α-WC nanopowders with increased adsorbed oxygen concentration is suggested to be one of the origins of increased creep activation energy $Q_{cr}$ in the testing of the UFG tungsten carbide samples. The $W_2C$ nanoparticles prevent the development of GBS in the central fine-grained parts of the samples and the motion of the lattice dislocations in the coarse-grained surface layers.

**Acknowledgments:** The study was supported by the Russian Foundation for Basic Research and Rosatom (Grant #20-33-90214). TEM investigations was carried out using the equipment of the Center Collective Use "Materials Science and Metallurgy" (National University of Science and Technology "MISIS") with the financial support of the Ministry of Science and Higher Education of the Russian Federation (Grant #075-15-2021-696).

**Author Contributions:** E.A. Lantcev – Investigation (SPS, density, hardness), Writing of manuscript; A.V. Nokhrin – Project administration, Funding acquisition, Conceptualization, Methodology, Formal analysis, Analysis of experimental results, Writing – review & editing, Visualization; V.N. Chuvil'deev – Formal analysis, Resources, Data curation, Supervision; M.S. Boldin – Investigation (SPS); Yu.V. Blagoveshchenskiy, N.V. Isaeva, and A.V. Terentyev – Investigation (DC arc plasma synthesis, annealing in hydrogen, chemical analysis, BET); P.V. Andreev and K.E. Smetanina – Investigation (XRD); A.A. Murashov – Investigation (SEM); N.Yu. Tabachkova – Investigation (TEM).

**Conflict of Interest:** The authors declare that they have no known competing financial interests or personal relationships that could have appeared to influence the work reported in this paper.

**Data Availability Statement:** Data available on request due to privacy restrictions.

**List of Figures**

Figure 1. Microstructure of the surface layer of Sample #2 after SPS. SEM

Figure 2. Results of XRD phase analysis of the tungsten carbide powders (the XRD curves marked with "P" index) add the ones of the ceramics (the XRD curves marked with "C" index)

Figure 3. Microstructure of the nanopowder #1: (a, b) – the nanopowder agglomerates at different magnifications (a – SEM, b – TEM), (c) – the amorphous layer on the nanopowder particle surfaces, (d) twins inside the tungsten carbide nanoparticles

Figure 4. Microstructure of the commercial powders #2 (a) and #3 (b). SEM

Figure 5. Temperature curves of the shrinkage L (dark markers) and of the shrinkage rate S (light markers): (a) powders with different initial particle sizes, (b) powders with different oxygen and carbon contents

Figure 6. Microstructure of ceramics #1 (a, b), #2 (c, d), and #3 (e, f) after SPS (a, c, e) and after creep tests at 1350 ºC (b, d, f). SEM

Figure 7. Microstructure of the ceramics #2.1 (a, b) and #2.2 (c, d) after SPS (a, c) and after the creep tests at 1350 ºC (b, d). SEM

Figure 8. Curves strain ($\varepsilon$) - holding time (t) for the ceramics #1 (a, b) and #2 (c, d) in the creep tests in Regime #1 (a, c) and in Regime #2 at 1325 °C (b, d)

Figure 9. Dependencies of the maximum strain $\varepsilon_{max}$ on the testing temperature in Regime #1. The line indices correspond to the ones of ceramics in Table 1

Figure 10. Photograph of the ceramic samples of Series #1 (a), #2 (b), and #3 (c) after the creep testing in Regime II: (1) initial sample, (2) 1300 °C (t = 30 min, σ = 70 MPa); (3) 1325 °C (t = 30 min, σ =70 MPa); (4) 1350 °C (t = 30 min, σ = 70 MPa). The samples were cut along the cylinder axis into two equal parts

Figure 11. Curves strain - time for the samples subjected to the mechanical treatment: (a) Regime #1, (b) Regime #2. Dark markers – ceramics #2.1 after standard treatment, light markers – ceramics #2.1 after additional mechanical treatment (ceramics #2.1$^*$)

Figure 12. Microstructure of samples of Series #2.1$^*$ after the creep tests at 1275 (a) and 1325 °C. SEM

Figure 13. Tungsten carbide samples #2.1 after the creep tests in Regime #1 (samples #2-4) and in Regime #2 (samples #5): #1 – initial sample, #2 – 1300 °C (t = 30 min, σ = 70 MPa); #3 – 1325 °C (t = 30 min, σ =70 MPa); #4 – 1350 °C (t = 30 min, σ = 70 MPa); #5 – 1325 °C (t = 50 min, stepwise increasing of the stress σ from 50 to 90 MPa). The samples were cut into two equal parts along the cylinder axis

Figure 14. Curves strain – testing time t for the ceramics #2.1 (a, b) and #2.2 (c, d) for the creep tests in Regime #1 at σ = 70 MPa (a, c) and in Regime #2 at 1325 °C (b, d)

Figure 15. Dependencies $\ln(\dot{\varepsilon})$ - $T_m/T$ (a) and $\ln(\dot{\varepsilon})$ – $\ln(\sigma/G)$ (b) for the fine-grained tungsten carbide samples #1-3

Figure 16. Dependence of the creep activation energy of tungsten carbide on the $W_2C$ mass fraction

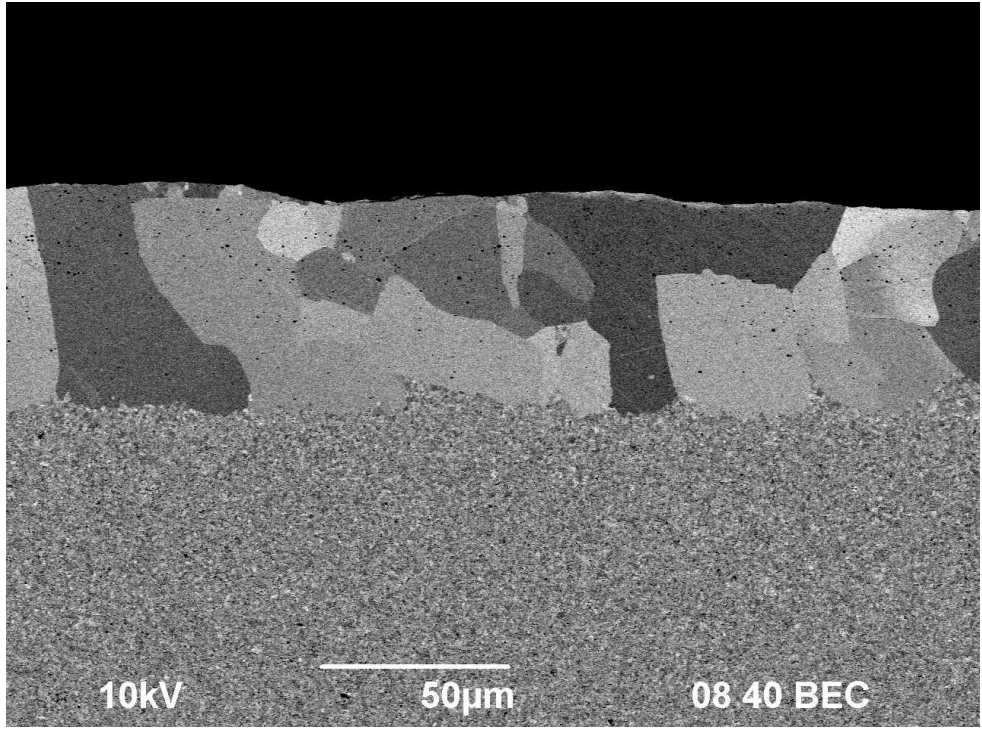

Figure 1

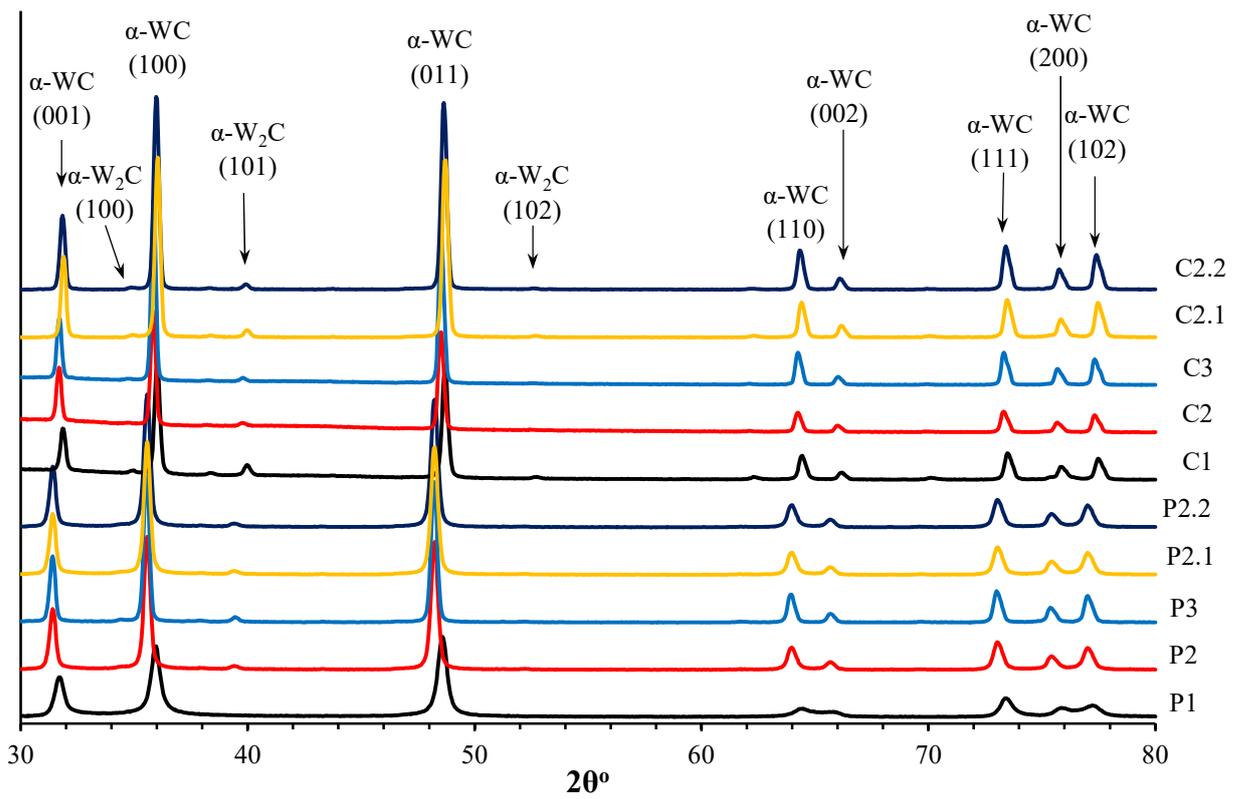

Figure 2

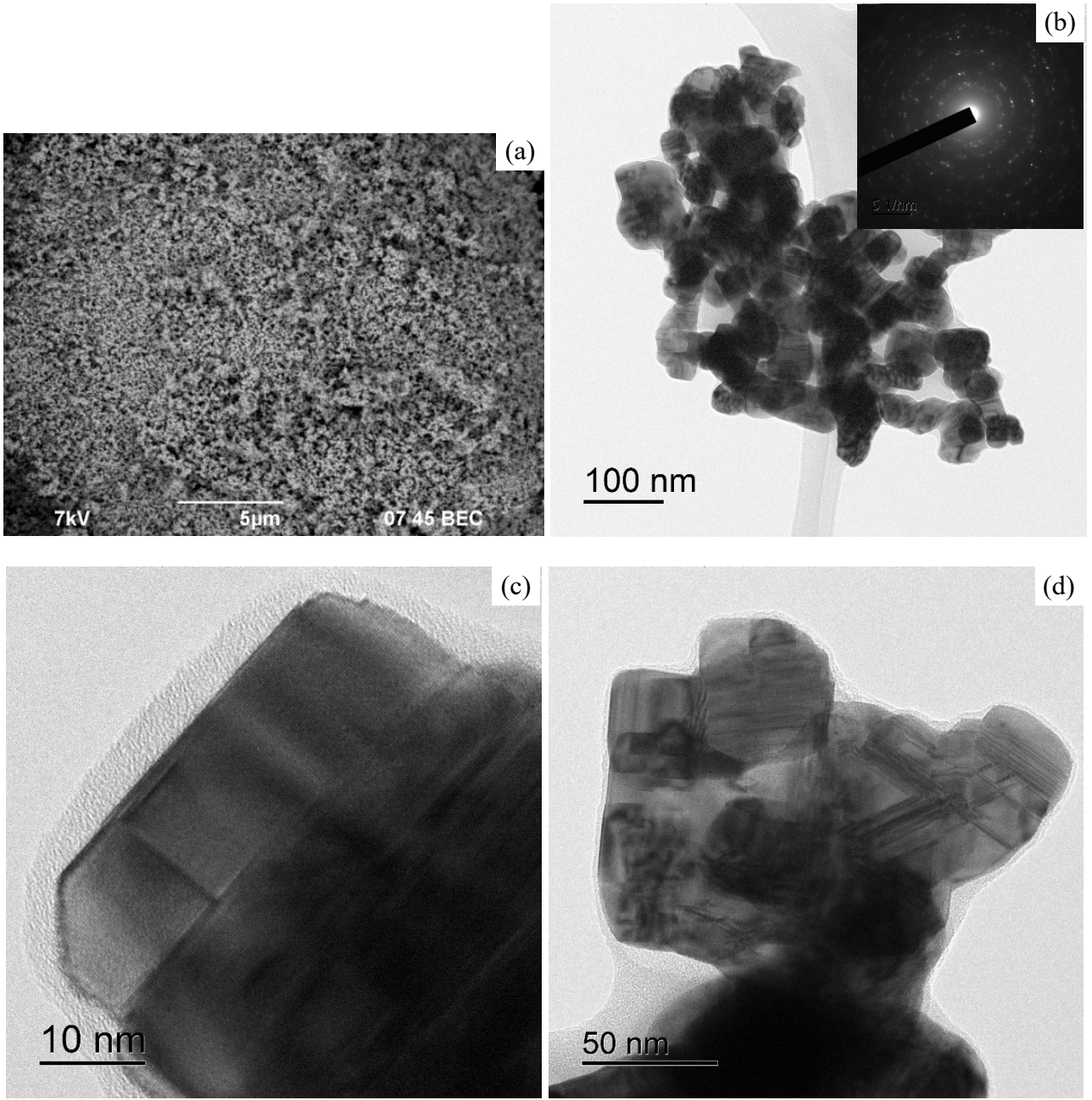

Figure 3

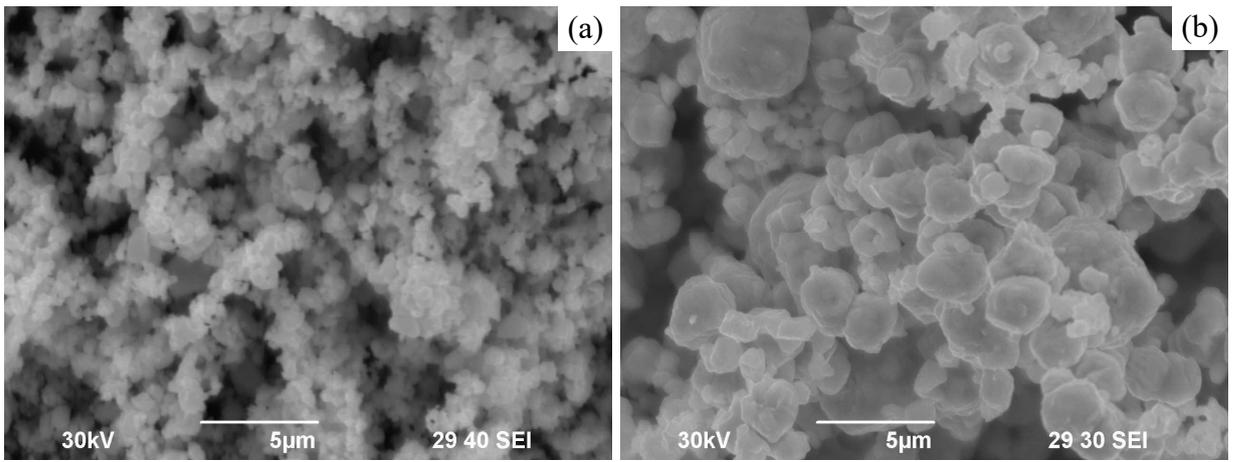

Figure 4

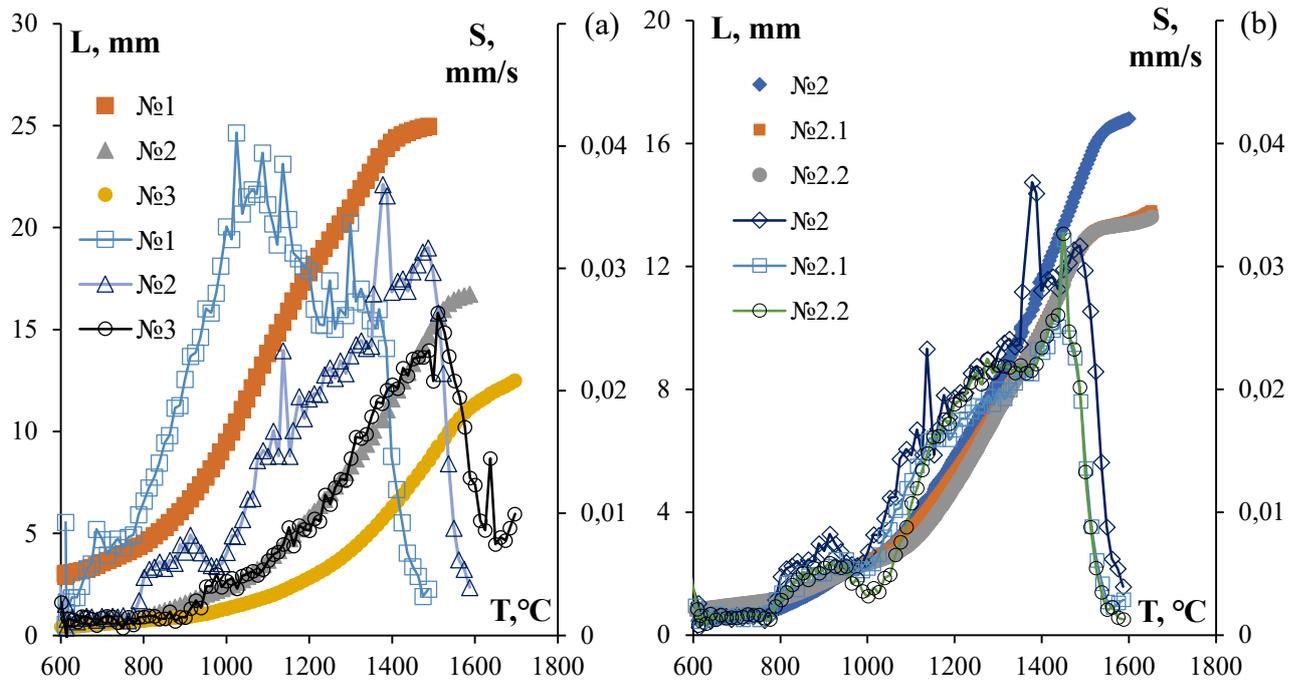

Figure 5

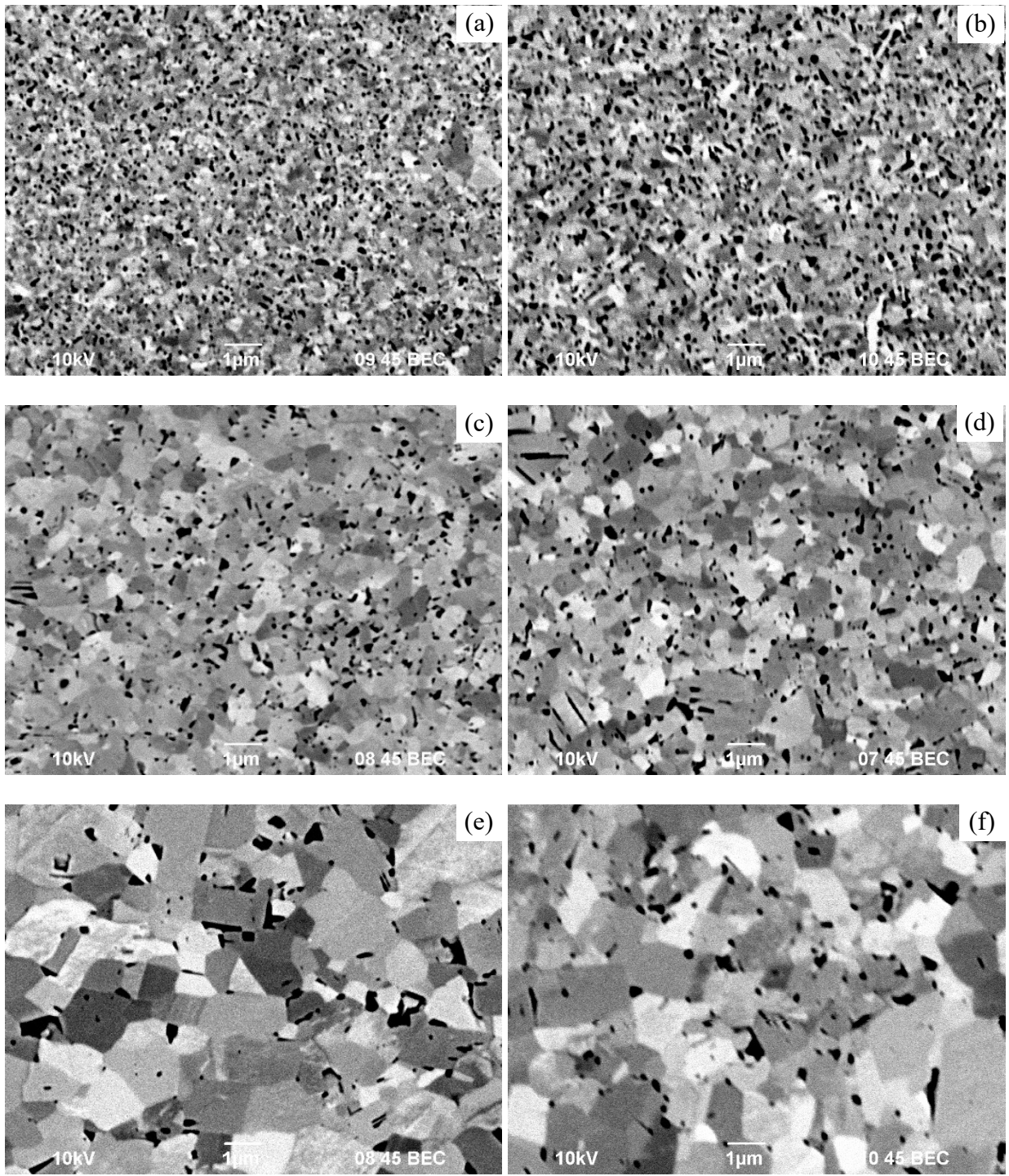

Figure 6

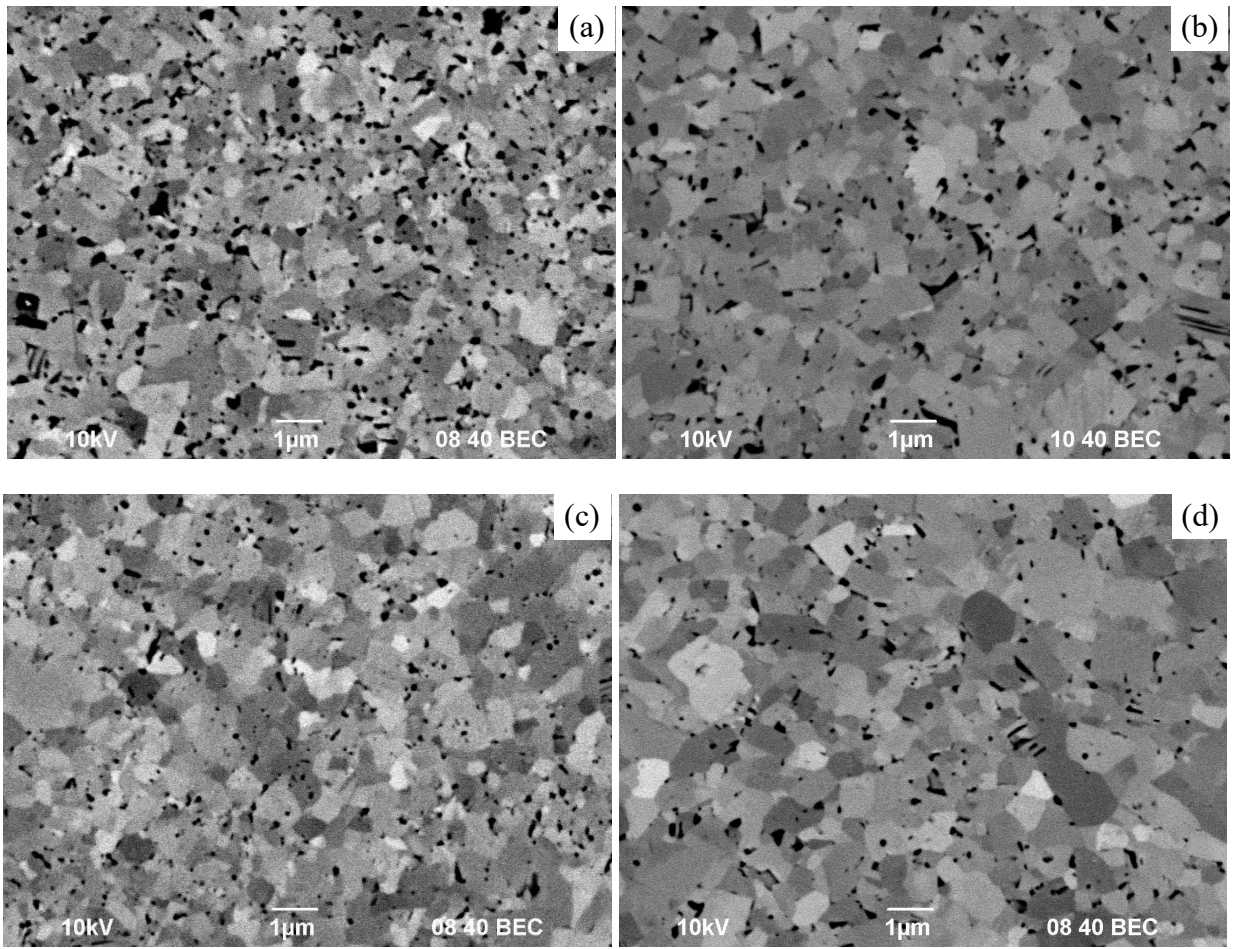

Figure 7

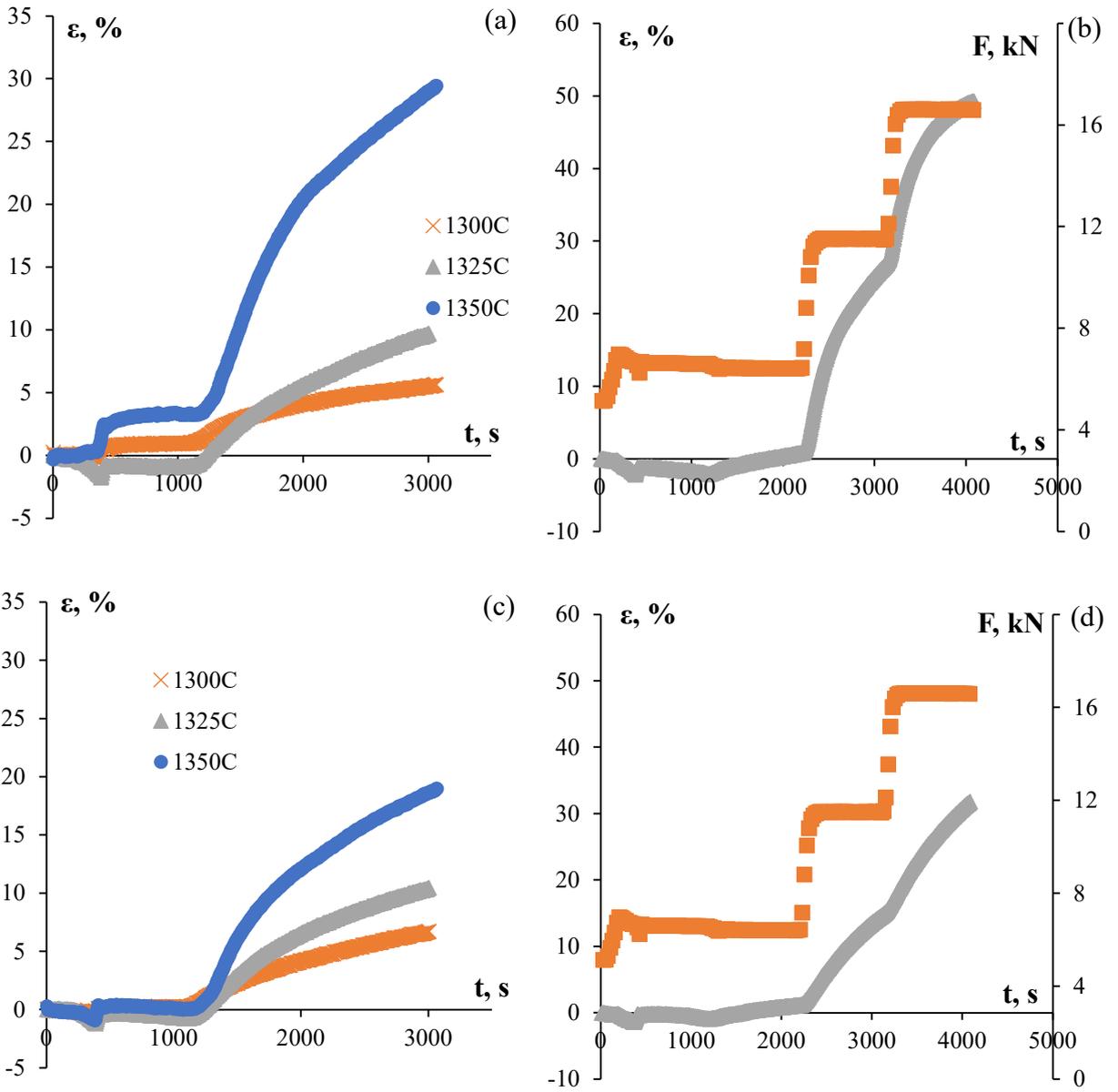

Figure 8

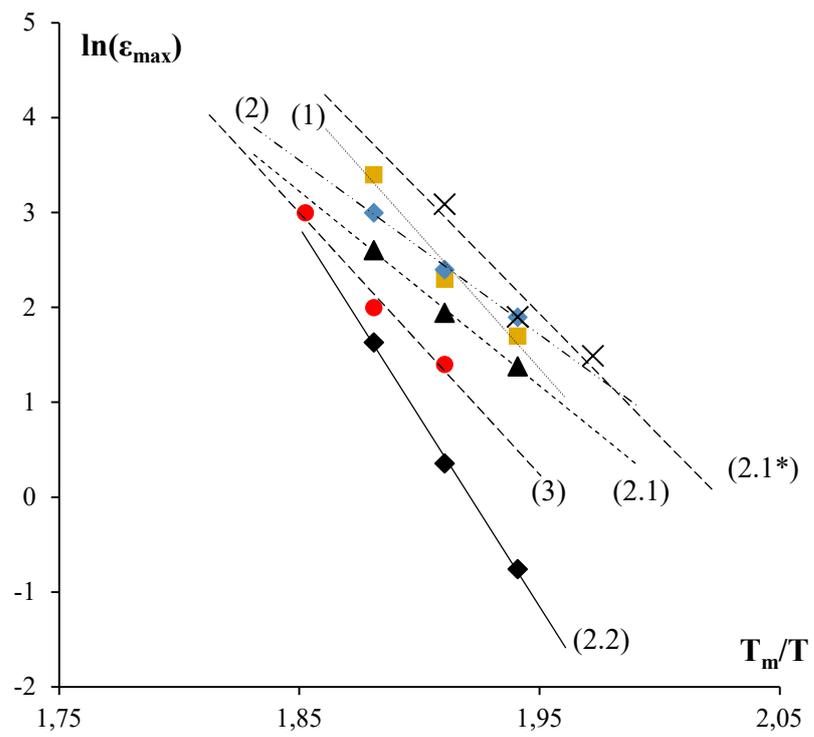

Figure 9

Figure 10

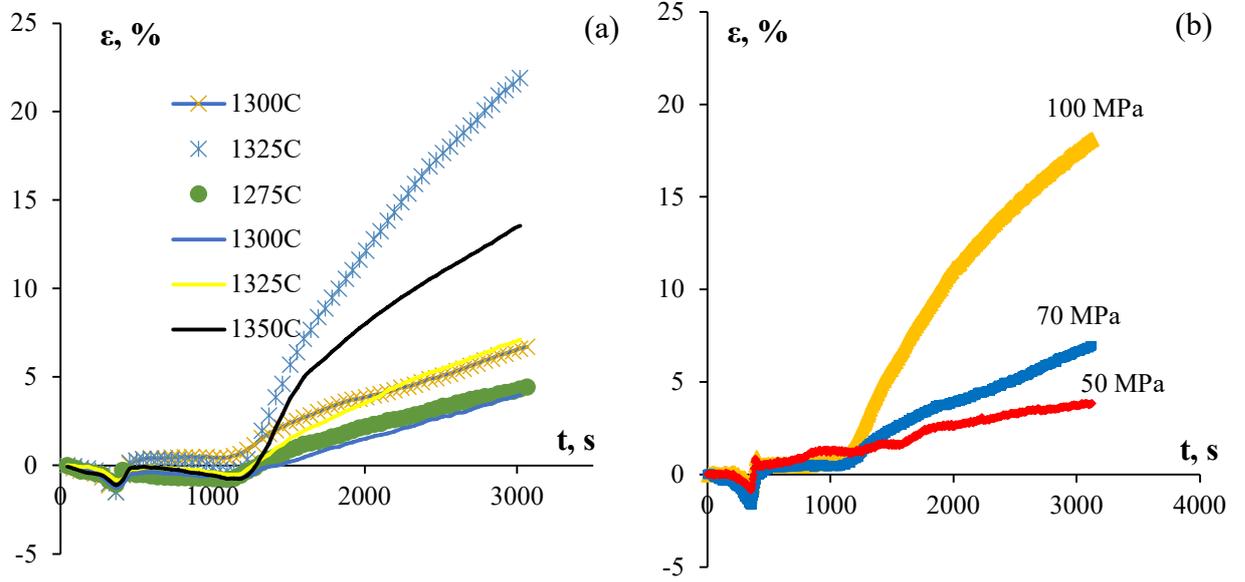

Figure 11

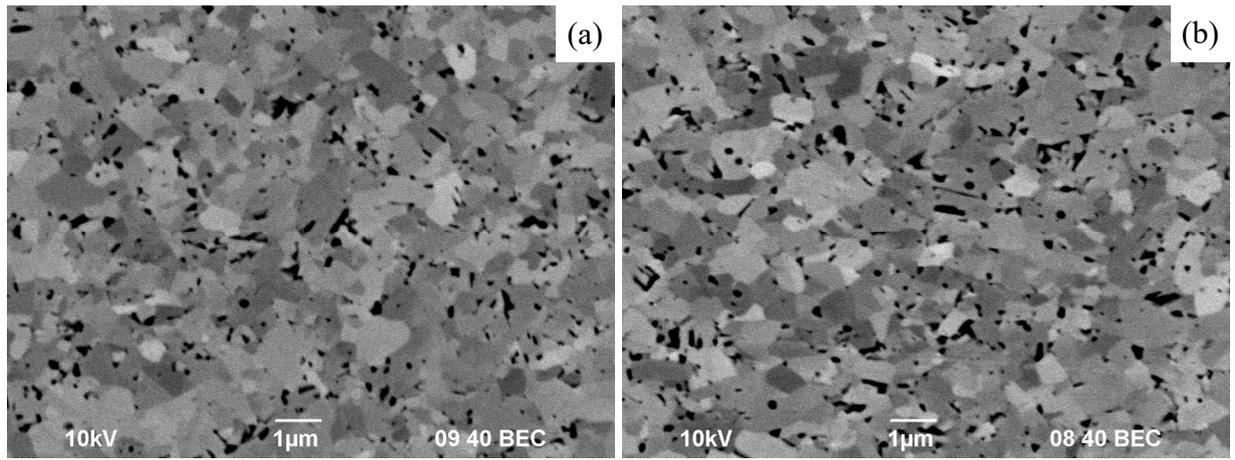

Figure 12

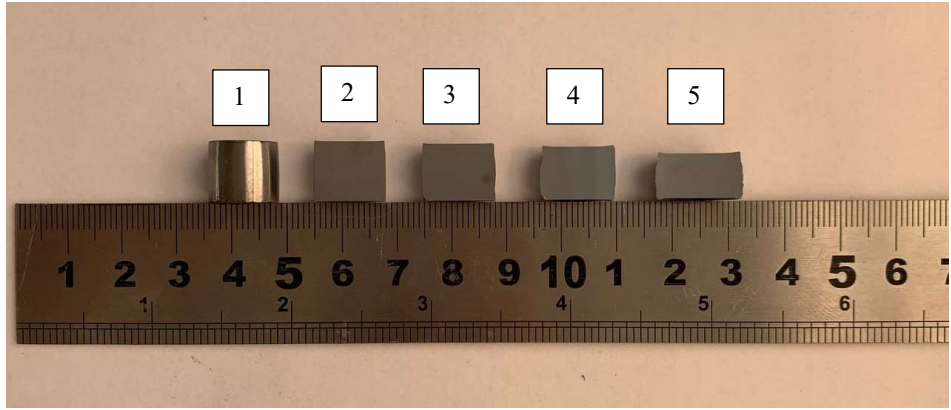

Figure 13

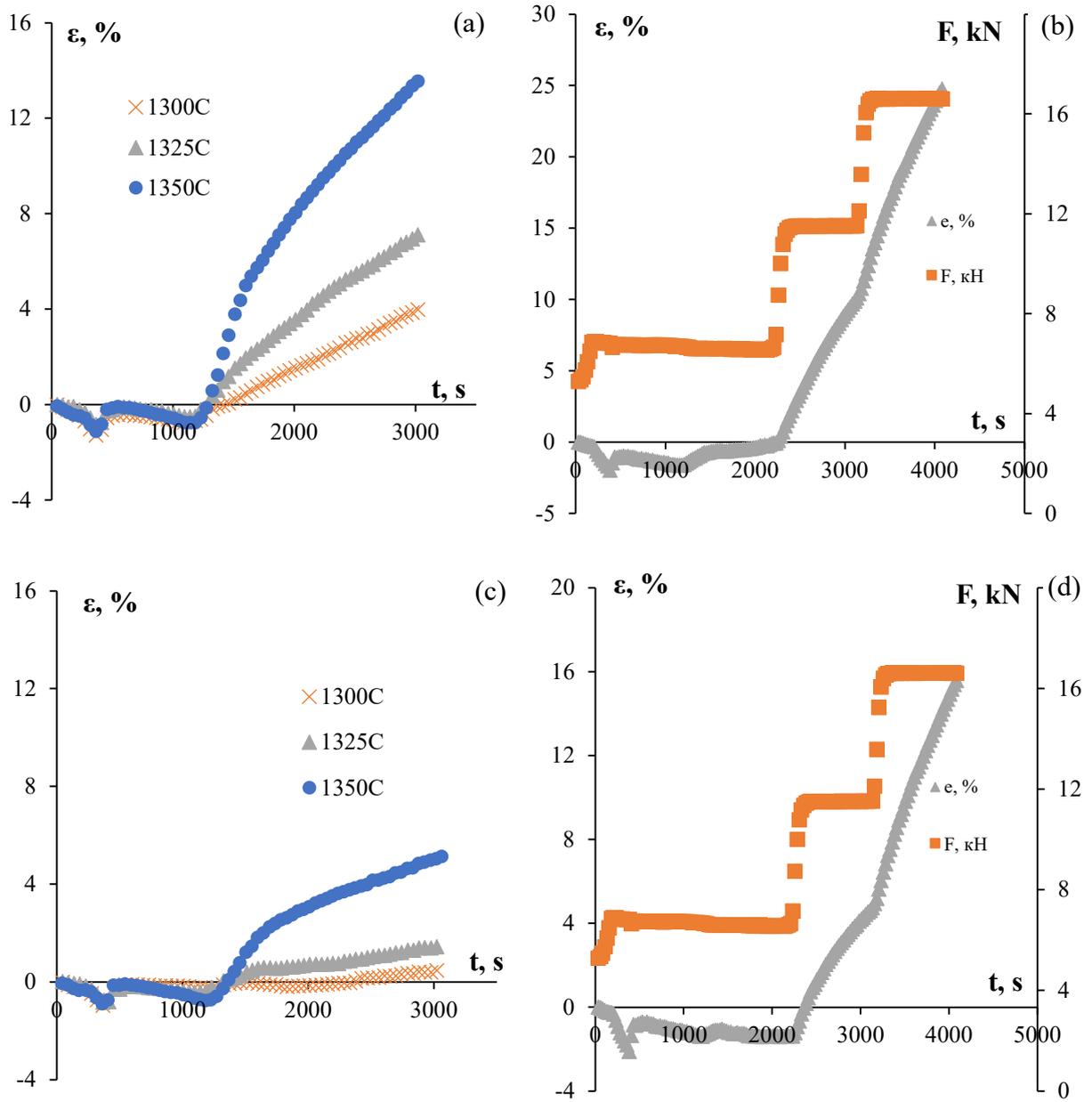

Figure 14

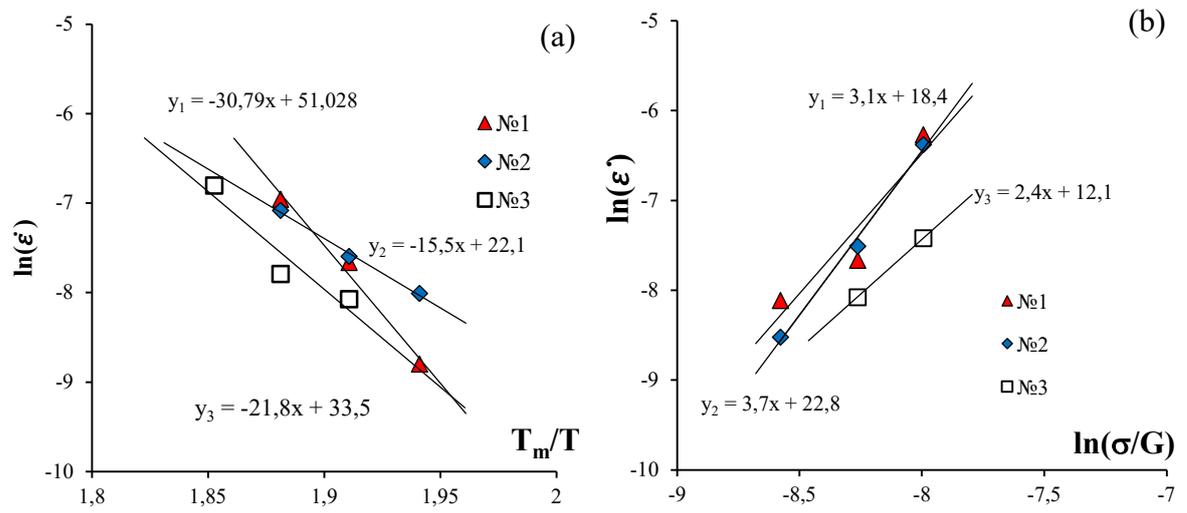

Figure 15

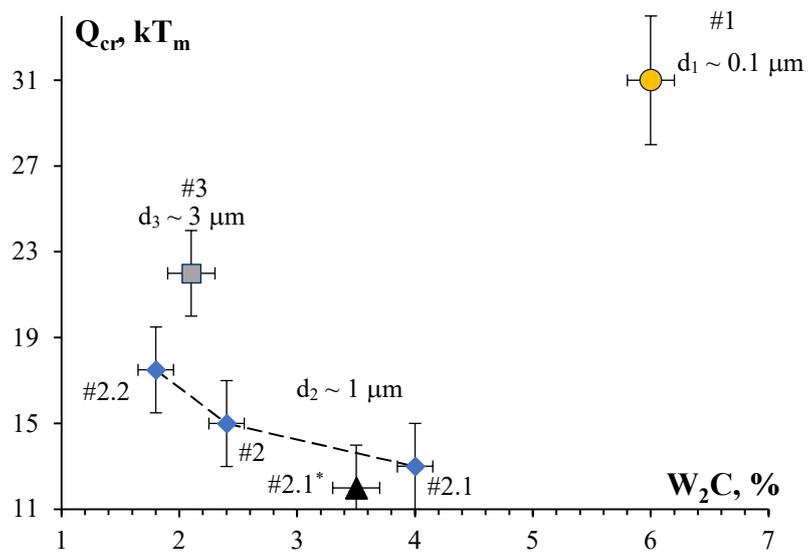

Figure 16